\documentclass[11pt]{article}
\usepackage{amssymb}
\def\Journal#1#2#3#4{{#1} {\bf #2}, #3 (#4)}

\def\CQG{\em Class. Quantum Grav.}
\def\JPA{\em J. Phys. A: Math. Gen.}
\def\PRD{\em Phys. Rev. D }
\def\GRG{\em Gen. Rel. Grav.}

\def\JMP{\em J. Math. Phys.}

\def\CMP{\em Commun. Math. Phys.}

\def\AIHP{\em Ann. Inst. H. Poincar\'e}

\def\CR{\em C.R. Acad. Sci. (Paris)}
\def\CP{\em Cahiers de Physique}
\def\NC{\em Nuovo Cimento}

\newcommand{\bm}[1]{\mbox{\boldmath $#1$}}
\def\d{{\rm d}}

\def\espaitemps{({\cal V},g)}
\def\varietat{{\cal V}}

\def\xiv{\vec \xi }
\def\etav{\vec \eta}
\def\lie{{\pounds}}
\def\xif{\mbox{\boldmath $ \xi $}}
\def\etaf{\mbox{\boldmath $ \eta $}}

\def\teta{* w}
\def\det{W}

\def\K{{\cal K}}

\def\CT#1#2{{\cal J}^{#2}\left(\vec #1\right)}
\def\CTf#1{{\bm{\cal J}}\left(\vec #1\right)}
\def\CB#1#2#3#4#5{j^{#5}_{#4}\left(\vec{#1},\vec{#2},\vec{#3}\right)}

\def\CBel#1#2#3{j^{#3}_{#2}\left(\vec{#1}\right)}
\def\CBf#1#2#3#4{\bm j^{#4}\left(\vec{#1},\vec{#2},\vec{#3}\right)}
\def\CBef#1#2#3{\bm j^{#3}\left(\vec{#1},\vec{#2}\right)}
\def\CBelf#1#2{\bm j^{#2}\left(\vec{#1}\right)}

\def\rr#1#2#3#4{r_{#4}\left(\vec{#1},\vec{#2},\vec{#3}\right)}
\def\rrf#1#2#3{\bm r\left(\vec{#1},\vec{#2},\vec{#3}\right)}

\def\ttc#1#2#3#4{t_{#4}\left(\vec{#1},\vec{#2},\vec{#3}\right)}
\def\ttf#1#2#3{\bm t\left(\vec{#1},\vec{#2},\vec{#3}\right)}

\def\be{\begin{equation}}
\def\ee{\end{equation}}
\def\bea{\begin{eqnarray}}
\def\eea{\end{eqnarray}}
\def\bean{\begin{eqnarray*}}
\def\eean{\end{eqnarray*}}

\def\proof{\noindent{\em Proof.\/}\hspace{3mm}}
\def\fin{\hfill \rule{2.5mm}{2.5mm}\\ \vspace{0mm}}

\newcounter{eqlletra}

\newtheorem{theorem}{Theorem}[section]
\newtheorem{result}{Result}[section]
\newtheorem{identity}{Identity}[section]

\def\be{\begin{equation}}
\def\ee{\end{equation}}
\def\bea{\begin{eqnarray}}
\def\eea{\end{eqnarray}}

\begin{document}
\setcounter{footnote}{2}
\title{Conserved superenergy currents}
\author{
Ruth Lazkoz$^1$\thanks{\tt wtblasar@lg.ehu.es}\,,
Jos\'e M.M. Senovilla$^1$\thanks{\tt wtpmasej@lg.ehu.es}\,,
\addtocounter{footnote}{0}%
and Ra\"{u}l Vera$^2$\thanks{\tt r.vera@qmul.ac.uk}
\\$^1$
{\sl F\'{\i}sica Te\'orica,
Universidad del Pa\'{\i}s Vasco,}\\
{\sl Apartado 644, 48080 Bilbao, Spain }
\\$^2${\sl School of Mathematical Sciences, Queen Mary,
University of London,}\\
{\sl Mile End Road, London E1 4NS, U.K. }
}
\date{\today}
\maketitle
\begin{abstract}
    We exploit once again the analogy between the energy-momentum 
tensor and the so-called ``superenergy'' tensors in order to build 
conserved currents in the presence of Killing vectors. First of 
all, we derive the divergence-free property of the gravitational 
superenergy currents under very general circumstances, even if the 
superenergy tensor is not divergence-free itself. The associated 
conserved quantities are explicitly computed for the 
Reissner-Nordstr\"om and Schwarzschild solutions. The remaining 
cases, when the above currents are not conserved, lead to the possibility
of an interchange of some superenergy quantities between the gravitational and
other physical fields in such a manner that the total, {\em mixed}, 
current may be conserved. Actually, this possibility has 
been recently proved to hold for the Einstein-Klein-Gordon system 
of field equations. By using an adequate family of known exact solutions, 
we present explicit and completely non-obvious
examples of such mixed conserved currents.
\end{abstract}

\section{Introduction and basic results}
Conserved quantities in Gravitation are either defined only at 
``infinity'', see e.g. \cite{WZ} and references therein, or they
arise as a consequence of the existence of di\-ver\-gence-free
vector fields, called {\it local currents},
and of the use of Gauss' theorem applied to appropriate domains of the
spacetime, see e.g. \cite{HE,MTW,SW}. A traditional way of constructing
divergence-free vector fields is by
using the intrinsic symmetries of the spacetime\footnote{A spacetime
$\espaitemps$ is a 4-dimensional manifold $\varietat$ with a metric
$g$ of Lorentzian signature (--,+,+,+). Indices in $\varietat$ run from 0 to
$3$ and are denoted by Greek small letters. Square and round brackets
denote the usual (anti-) symmetrisation of indices. The tensor and exterior
products are denoted by $\otimes$ and $\wedge$, respectively, boldface letters
are used for 1-forms and arrowed symbols for vectors,
and the exterior differential is denoted by $d$. The Lie derivative
with respect to the vector field $\xiv$ is written as $\lie_{\xiv}$,
and the covariant derivative by $\nabla$. Equalities by definition are denoted by
$\equiv$, and the end of a proof is signalled
by \rule{2.5mm}{2.5mm}.}: if $\xiv$ is a
Killing vector field, $\nabla_{(\alpha}\xi_{\beta)}=0$, representing
an infinitesimal isometry, then many local currents and associated
conserved quantities arise, such as for instance every vector field of
the form $A\xiv$, where $A$ is any function satisfying $\xiv (A) =0$.
This is a simple consequence of the invariance of the geometrical
background under the Killing vector field. Of course, if there are
several Killing vectors $\xiv_{L}$ in the spacetime, then any linear
combination of the vector fields of type $A^{L}\xiv _{L}$ is also
divergence-free whenever each $A^{L}$ is invariant with respect to
the corresponding $\xiv_{L}$.

There are also other, physically more interesting,
types of locally conserved currents constructible from $\xiv$
together with other objects. Take, for example, the typical case of
$t^{\mu\nu}\xi_{\nu}$ for a {\it divergence-free symmetric} tensor
$t^{\mu\nu}$. Obviously we have
$$
\nabla_{\mu}\left(t^{\mu\nu}\xi_{\nu}\right)=
\left(\nabla_{\mu}t^{\mu\nu}\right)\xi_{\nu} +
t^{[\mu\nu]}\nabla_{\mu}\xi_{\nu}=0
$$
where the Killing property $\nabla_{(\alpha}\xi_{\beta)}=0$ has been used.
Observe that both the divergence-free property {\it and} the symmetry
of $t^{\mu\nu}$ are needed here. In fact the divergencelessness of
$t^{\mu\nu}$ is a sufficient but not necessary condition,
as one only needs that
$\xiv$ and $\nabla\cdot t$ be orthogonal for the first term to vanish.
Furthermore, observe that just a {\it conformal}
Killing vector field ($\nabla_{(\alpha}\xi_{\beta)}\propto g_{\alpha\beta}$)
is required if $t_{\mu\nu}$ is also traceless.

A particular important example of this type of locally conserved currents is
provided by the case of the total energy-momentum tensor $T^{\mu\nu}$
of the matter fields in General Relativity. In this case, due to the
Einstein field equations, $T^{\mu\nu}$ is symmetric and divergenceless, and
thus any Killing vector field $\xiv$ provides a divergence-free current
defined by
\be
\CT{\xi}{\alpha}\equiv T^{\alpha\beta}\xi_{\beta} .
\label{eq:ct}
\ee
These currents have also a physical meaning which depends on the particular
character of the Killing vector $\xiv$. For instance, in stationary spacetimes
there is a timelike Killing vector
$\xiv$, and the corresponding current $\vec\CT{\xi}{}$
can be thought of as the local energy-momentum vector of
 the matter in the stationary system of reference.
Similarly, for translational or rotational Killing vectors one obtains
currents representing the whole linear or angular momentum, respectively,
at least in flat spacetimes.

Two important remarks are in order here. First, in many cases the mentioned
currents $\vec\CT{\xi}{}$ are of the form referred to above, that is,
$\vec\CT{\xi}{}= B^{L}\xiv _{L}$ for some particular functions $B^L$.
This does not deprive $\vec\CT{\xi}{}$
of its physical significance. Rather one has to see this result as stating
that, among the huge variety of divergence-free vectors of the form
$A^{L}\xiv _{L}$ with general $A^L$, the particular one with the functions
$A^L =B^L$ have a specially relevant physical meaning. Several
cases in which this happens are given by the following known result
\cite{kundttrump,papapetrou,carter69,chuscomm,KRAM,raultesis}.\footnote{Recall
that a vector field
$\vec v$ is called  hypersurface-orthogonal (also
``integrable'') if $\bm v \wedge \d \bm v=0$, and that
two vector fields orthogonal to
the non-null surfaces spanned by
two given vector fields $\vec v$ and $\vec w$
generate surfaces whenever the
two 1-forms $\bm v$ and $\bm w$ satisfy
$\bm v \wedge \bm w \wedge \d \bm w=\bm v \wedge \bm w \wedge \d \bm v=0$.
Thus a $G_2$-group of motions generated by two Killing vectors $\vec\xi$ and
$\vec\eta$ is said to be acting orthogonally transitively
on non-null orbits when
there exists a family of surfaces orthogonal to the orbits,
which means that $\bm \xi \wedge \bm \eta
\wedge \d \bm \eta= \bm \xi \wedge \bm \eta \wedge \d \bm \xi=0$.}
\begin{result}
Assume Einstein's field equations of General Relativity hold.
\begin{enumerate}
\item If $\vec \xi$ is a hypersurface-orthogonal Killing vector, then
$\CTf{\xi}\wedge \xif =0$.
\item If $\xiv$ and $\vec{\eta}$ are two commuting Killing
vectors $[\vec \xi,\vec \eta]=0$ and they act
orthogonally transitively on non-null surfaces, then
$\CTf{\xi}\wedge \xif \wedge \etaf=
\CTf{\eta}\wedge \xif \wedge \etaf =0$.
\end{enumerate}
\label{teo:orthtran}
\end{result}
The first of these results means that if there are (locally) hypersurfaces
orthogonal to the Killing vector, then the corresponding current lies in
the direction of the Killing vector itself. This implies that the integration
of $\vec\CT{\xi}{}\cdot \xiv$ (and of $\vec\CT{\xi}{}\cdot \xiv /|\xiv|$)
over any two different hypersurfaces orthogonal to $\xiv$ is the same.
The second one states that any two
(group-forming) Killing vectors whose surfaces of transitivity are orthogonal
to a family of surfaces have currents which lie necessarily in the
orbits of the $G_2$ group generated by them. Of course, both results
are quite natural and follow from the invertibility of
the orthogonally transitive Abelian $G_2$ group
(which reduces to hypersurface-orthogonality for 1-dim groups) \cite{carter69}.

The second remark concerns the case when there are several types of matter
fields in the spacetime, and the corresponding energy-momentum tensors.
For simplicity, assume that there are two different sorts of matter
contents (an electromagnetic field and a fluid for instance), with
corresponding energy-momentum tensors $T^{(1)}_{\alpha\beta}$ and
$T^{(2)}_{\alpha\beta}$, forming an isolated system. Then, only the {\it total}
energy-momentum tensor $T_{\alpha\beta}\equiv T^{(1)}_{\alpha\beta}+
T^{(2)}_{\alpha\beta}$ is divergence-free, and therefore the partial
currents constructed from $T^{(1)}_{\alpha\beta}$ or
$T^{(2)}_{\alpha\beta}$ are not conserved separately in general.
Of course, this is physically
meaningful and reasonable, as it leads to the conservation of the
corresponding total quantity, and thereby implies the {\it exchange}
of some energy-momentum properties  between the two fields
involved. A classical example of this situation can be found in the standard
reference \cite{LL} for the case of Special Relativity, where one can see
that if the sources of the electromagnetic field are taken into account
then the energy-momentum tensor of the electromagnetic field is not
divergence-free, but its combination with the energy-momentum tensor
of the charges creating the field provides a total energy-momentum
tensor which is divergence-free and thereby provides the usual conserved
currents associated to the ten Killing vectors of flat spacetime. This proves
in particular that the energy-momentum properties can be transferred from
one field to another, or from matter to the fields, and vice versa.
This transference property together with the positivity and
the conservation are the basis for the paramount importance of the
concept of energy-momentum in physical theories.

In this paper we shall analyse other locally conserved currents arising
in Gravitation which keep the mentioned three properties: positivity,
local conservation, and interchange. These local currents arise due to the
existence of the so-called ``superenergy'' tensors,
see \cite{B1,B4,Beltesis,B2,Rob,S,supertotxo} and references therein,
and, as we shall prove, they
have analogous attributes to those so far mentioned for the
energy-momentum currents (\ref{eq:ct}). The type of currents we shall 
analyse have expressions such as
$$
T^{\alpha\beta\lambda\mu}\,\xi_{\beta}\xi_{\lambda}\xi_{\mu},
$$
or adequate generalisations of this using different Killing vectors, where 
$T^{\alpha\beta\lambda\mu}$ will be any of the available superenergy tensors 
for the physical fields involved.

In particular, we start by
constructing the corresponding currents for the Bel superenergy (s-e) tensor
\cite{B4,Beltesis,supertotxo}, which is divergence-free in vacuum (where it is
called the Bel-Robinson tensor) and in Einstein spaces, and, therefore, it
can lead to conserved s-e currents.
We shall see in Section \ref{Belcurrents}
that, in fact, all the currents constructed from the Bel tensor (in
vacuum or not) and a Killing vector satisfy properties analogous to
those of Result \ref{teo:orthtran}, which will result in
divergence-free currents in many general situations. Some specific
examples of these currents are written down and analysed in Section
\ref{R-N} leading to physical consequences of some interest.

Nevertheless, in completely general situations, the Bel s-e currents
will not be divergence-free; something which seems natural if the s-e
concept is to have any physical meaning at all. This is because if in the
spacetime  other fields than the gravitational one are present, one should expect in principle that the conserved current
be formed by a combination of the sum of the gravitational Bel s-e
current and the s-e current of the matter fields involved, in analogy
with the energy-momentum currents, as explained
before. This is supported also by the fact that the s-e currents of
the matter fields are conserved in {\em flat} spacetime --that is, in
the absence of gravitational field--, see for a proof
\cite{S2,supertotxo,Tey}. As a matter of fact, the existence of such
{\em mixed} divergence-free currents has been rigorously proved in
\cite{S2,supertotxo} for the case of a minimally coupled scalar field,
that is, for the Einstein-Klein-Gordon situation, whenever there is a
Killing vector. As explicitly shown in \cite{S2,supertotxo}, none
of the two single currents, the scalar field s-e current nor the Bel
one, are divergence-free separately in general. These results are
summarised briefly in Section \ref{mixed}.

One of the purposes of this paper is to compute these divergence-free
mixed currents in some particular situations, thereby providing
explicit examples of the exchange of s-e between the scalar and
gravitational fields, and the conservation of combined
quantities. However, there arise several technical difficulties. First
of all, the result for the Bel tensor which is analogous to Result
\ref{teo:orthtran} above, presented in Results \ref{teo:belcurrents1}
and \ref{teo:belcurrents2} of Section \ref{Belcurrents}, implies that
(see Theorem \ref{key}): 1) if the Killing vector is
hypersurface-orthogonal, then its corresponding Bel current is
automatically divergence-free; and 2) if two Killing vectors form an
Abelian group which acts orthogonally transitively on non-null
surfaces, then all the corresponding Bel currents are divergence-free
too. Thus, even though these results are physically interesting on
their own, they prevent the existence of {\em mixed} currents in those
cases. Hence, we have to resort to other more general cases in order
to find such mixed currents. But then a second problem arises, namely,
to find explicit spacetimes which are solution of the
Einstein-Klein-Gordon equations under the needed more general
circumstances: there are hardly any known solutions with a
non-hypersurface-orthogonal Killing vector, and very few with a
non-orthogonally transitive 2-parameter group.

The good news is, though, that there is at least an explicitly known
family of solutions of the type we need. These solutions were found by
Wils in \cite{Wils} and we will use them in the second part of Section
\ref{mixed} to construct the sought {\em mixed} divergence-free s-e
currents.  This will provide explicit examples of the s-e
interchange and of conserved currents which would have been very 
difficult to find, and even more to be singled out, if the concept of 
superenergy tensors had not been available.

\section{General results on Bel currents}
\label{Belcurrents}
We will use the notation and terminology of \cite{supertotxo} for the
superenergy tensors. The Bel tensor, introduced by Bel more than 40
years ago \cite{B4,Beltesis}, is the basic superenergy tensor of the
total gravitational field, that is to say, the s-e tensor 
$T_{\alpha\beta\lambda\mu}\left\{R_{[2],[2]}\right\}$ associated
to the Riemann tensor $R_{\alpha\beta\lambda\mu}$. Its explicit
expression reads \cite{Ch,BoS,supertotxo}
\begin{eqnarray}
B_{\alpha\beta\lambda\mu}\equiv
R_{\alpha\rho\lambda\sigma}
R_{\beta}{}^{\rho}{}_{\mu}{}^{\sigma}
+R_{\alpha\rho\mu\sigma}
R_{\beta}{}^{\rho}{}_{\lambda}{}^{\sigma}- \hspace{1cm} \nonumber\\
-\frac{1}{2}g_{\alpha\beta}
R_{\rho\tau\lambda\sigma}R^{\rho\tau}{}_{\mu}{}^{\sigma}
-\frac{1}{2}g_{\lambda\mu}
R_{\alpha\rho\sigma\tau}R_{\beta}{}^{\rho\sigma\tau}+
\frac{1}{8}g_{\alpha\beta}g_{\lambda\mu}
R_{\rho\tau\sigma\nu}
R^{\rho\tau\sigma\nu}, \label{Bel}
\end{eqnarray}
from where the following properties explicitly arise:
\begin{equation}
B_{\alpha\beta\lambda\mu}=B_{(\alpha\beta)(\lambda\mu)}=
B_{\lambda\mu\alpha\beta},\hspace{2mm}
B^{\rho}{}_{\rho\lambda\mu}=0.\label{belsym}
\end{equation}

Using the second Bianchi identity $\nabla_{[\nu}R_{\alpha\beta]\lambda\mu}=0$
one obtains the following expression for the divergence of the Bel tensor:
\begin{equation}
\nabla_{\alpha}B^{\alpha\beta\lambda\mu}=
R^{\beta\hspace{1mm}\lambda}_{\hspace{1mm}\rho\hspace{2mm}\sigma}
J^{\mu\sigma\rho}+R^{\beta\hspace{1mm}\mu}_{\hspace{1mm}\rho\hspace{2mm}\sigma}
J^{\lambda\sigma\rho}-\frac{1}{2}g^{\lambda\mu}
R^{\beta}_{\hspace{1mm}\rho\sigma\gamma}J^{\sigma\gamma\rho},\label{divbel}
\end{equation}
where $J_{\lambda\mu\beta}=-J_{\mu\lambda\beta}\equiv
\nabla_{\lambda}R_{\mu\beta}-\nabla_{\mu}R_{\lambda\beta}$ and $R_{\beta\mu}$
is the Ricci tensor.
Note that because of (\ref{belsym}) this is the only independent
divergence of the Bel tensor. From (\ref{divbel}) we obtain the fundamental
result that $B$ is divergence-free when $J_{\lambda\mu\beta}=0$.
This happens, for instance, in all Einstein spaces (where
$R_{\mu\nu}=\Lambda g_{\mu\nu}$), so that
in particular this implies that the Bel-Robinson tensor
\cite{B1,Beltesis,B2,Rob}, which is the s-e tensor of the Weyl 
conformal tensor and coincides with $B$ in Ricci-flat spacetimes \cite{BoS},
is divergence-free.
The divergence-free property of the Bel or Bel-Robinson tensors
in these cases allows for the construction of divergence-free currents whenever
there is a Killing vector. 

These currents are built in a similar way to those formed with the energy-momentum
tensor, as we show next. Following \cite{supertotxo} one can define the
{\em Bel current} with respect to any three Killing vector fields
$\vec \xi^{}_1$, $\vec \xi^{}_2$, $\vec \xi^{}_3$ (they do not need 
to be different!) as
\begin{equation}
  \label{eq:current}
  j_{\mu}\left(\vec{\xi}^{}_1,\vec{\xi}^{}_2,\vec{\xi}^{}_3\right)\equiv
B_{(\alpha\beta\lambda)\mu}\,\xi^{\alpha}_1\xi^{\beta}_2\xi^{\lambda}_3=
B_{(\alpha\beta\lambda\mu)}\,\xi^{\alpha}_1\xi^{\beta}_2\xi^{\lambda}_3.
\end{equation}
To avoid unnecessary writing we will
omit the repetition of the Killing vectors:
$\CBef{\xi_1}{\xi_2}{}\equiv \CBf{\xi_1}{\xi_2}{\xi_2}{}$
and $\CBelf{\xi}{}\equiv \CBf{\xi}{\xi}{\xi}{}$.
The divergence of these currents can be easily computed to give
{\setlength\arraycolsep{0.1pt}
\begin{eqnarray*}
\nabla_{\mu}\CB{\xi_1}{\xi_2}{\xi_3}{}{\mu}&=&
\nabla^\mu B_{(\alpha\beta\lambda\mu)}
\,\xi^{\alpha}_1\xi^{\beta}_2\xi^{\lambda}_3 +
3 B_{(\alpha\beta\lambda\mu)}
\nabla^{(\mu}_{ } \xi^{\alpha)}_{(1}\xi^{\beta}_2\xi^{\lambda}_{3)}
=\nabla^\mu B_{(\alpha\beta\lambda\mu)}
\,\xi^{\alpha}_1\xi^{\beta}_2\xi^{\lambda}_3,
\end{eqnarray*}}
where we have used the fact that the $\vec\xi_a$ are Killing vectors:
$\nabla^{(\alpha} \xi^{\beta)}_a=0$ ($a=1,2,3$.)
Therefore, the vanishing of the divergence of the Bel tensor
implies the vanishing of that of
$\vec{j}\left(\vec\xi_1,\vec\xi_2,\vec\xi_3\right)$,
defining, thereby, a conserved current.\footnote{Observe that
{\em only} in strictly Ricci-flat 
spacetimes we can use conformal Killing vectors for
$\vec \xi^{}_1$, $\vec \xi^{}_2$, $\vec \xi^{}_3$ obtaining also  
local currents of type (\ref{eq:current}), as only in that case the Bel-Robinson
tensor is divergence- {\em and} trace-free. Then, one can involve conformal 
Killing tensors---or even conformal Yano-Killing tensors as in \cite{J}--.}

Nevertheless, the Bel currents are {\em not only} conserved when the Bel tensor is
divergence-free.
Following the general arguments in \cite{carter69} showing the invertibility
of orthogonal transitive Abelian $G_2$ groups (and, in particular,
integrable 1-dim groups) acting on non-null orbits,
since the Bel tensor is intrinsically defined, if the metric admits
an invertible group
then the Bel tensor must be invertible implying that
its contraction with any odd number of 
Killing vectors provides a tensor tangent to the group orbits.
Therefore, the s-e currents constructed from a hypersurface-orthogonal
Killing vector or two non-null Killing vectors
generating an Abelian orthogonally-transitive
$G_2$ will necessarilly be tangent to the orbits of the corresponding groups.
This is stated as follows:
\begin{result}
If $\vec\xi$ is a hypersurface-orthogonal Killing vector, then $
\CBelf{\xi}{}\!\wedge\bm{\xi}=0$.
\label{teo:belcurrents1}
\end{result}
\begin{result}
Let $\vec \xi$ and $\vec \eta$ be two independent commuting
Killing vector fields spanning non-null surfaces.
Then, if they act orthogonally transitively,
their four associated Bel currents satisfy
\be
\CBelf{\xi}{}\!\wedge \bm{\xi}\wedge\bm{\eta}=
\CBelf{\eta}{}\!\wedge \bm{\xi}\wedge\bm{\eta}=
\CBef{\xi}{\eta}{}\! \wedge \bm{\xi}\wedge\bm{\eta}=
\CBef{\eta}{\xi}{}\! \wedge \bm{\xi}\wedge\bm{\eta}=0.
\label{4}
\ee
\label{teo:belcurrents2}
\end{result}
These currents satisfy then properties completely analogous to that of the
energy-momentum currents stated in Result \ref{teo:orthtran},
and therefore they will be
divergence-free in quite general cases, as we will show below.
For the sake of completeness, we provide
the identities involving the s-e currents and the geometrical
properties of 1-dim and 2-dim groups of isometries,
from which Results \ref{teo:belcurrents1} and \ref{teo:belcurrents2}
trivially follow.

\begin{identity}
Let $\vec\xi$ be a Killing vector field in $\espaitemps$. Then we have
\be
\xi^\alpha \xi^\beta \xi^\lambda B_{\alpha \beta\lambda [\mu} \xi_{\gamma]}=
\frac{3}{2}\left\{ \xi_\lambda \xi_\beta R^{\lambda\sigma\beta\tau}
\nabla_\tau \left( \xi_{[\mu}\xi_{\gamma;\sigma]}\right)+
\xi^\beta {R^{\rho\tau\sigma}}_{[\mu} \xi_{\gamma]} \nabla_\sigma
\left(\xi_{[\rho} \xi_{\beta;\tau]} \right) \right\}.
\label{eq:xixixixi}
\ee
\label{teo:xixixixi}
\end{identity}
\proof
Since $\vec\xi$ is a Killing vector we have \cite{Eis,MTW}
\be
\nabla_{\alpha}\nabla_{\beta}\xi_{\lambda}=
\xi_{\rho}R^{\rho}{}_{\alpha\beta\lambda},
\label{eq:basic}
\ee
so that direct contraction of the Bel tensor (\ref{Bel}) with $\vec \xi$ yields,
on using (\ref{eq:basic}),
{\setlength\arraycolsep{0.1pt}
\begin{eqnarray*}
\xi^\alpha \xi^\beta \xi^\lambda B_{\alpha\beta\lambda\mu}&=&
2\xi^\lambda
\nabla_\rho \nabla_\sigma \xi_\lambda \nabla^\rho \nabla^\sigma \xi_\mu
+\frac{1}{2}\xi^\lambda \xi_\lambda \nabla_\sigma \nabla_\rho \xi_\tau
R^{\rho\tau\sigma}{}_{\mu}-\\
&-&\frac{1}{2}\xi_\mu \nabla_\rho \nabla_\sigma \xi_\lambda
\nabla^\rho \nabla^\sigma \xi^\lambda+
\frac{1}{8}\xi^\lambda \xi_\lambda \xi_\mu
R^{\rho\sigma\tau\nu}R_{\rho\sigma\tau\nu}.
\end{eqnarray*}}
The exterior product of this one-form with $\bm\xi$ gives then
\begin{equation}
  \label{eq:jotaxi}
  \xi^\alpha \xi^\beta \xi^\lambda B_{\alpha\beta\lambda[\mu}\xi_{\gamma]}=
2\xi_\lambda
\nabla^\rho \nabla^\sigma \xi^\lambda \nabla_\rho \nabla_\sigma \xi_{[\mu}
\xi_{\gamma]}+\frac{1}{2}\xi^\lambda \xi_\lambda
\nabla_\sigma \nabla_\rho \xi_\tau
R^{\rho\tau\sigma}{}_{[\mu}\xi_{\gamma]}.
\end{equation}
Expanding the covariant derivative of the three-form
$\bm\xi\wedge \d \bm\xi$ we get
\begin{equation}
  \label{eq:1-se2}
  3\;\nabla_\rho\left( \xi_{[\gamma}\xi_{\sigma;\mu]}\right)=
-3\;\nabla_\gamma\xi_{[\rho}\nabla_\mu\xi_{\sigma]}+
\xi_\sigma\nabla_\rho \nabla_\gamma \xi_\mu-
2\;\nabla_\rho \nabla_\sigma \xi_{[\mu}\xi_{\gamma]}.
\end{equation}
and taking into account that
$\xi_\lambda\nabla^\rho \nabla^\sigma \xi^\lambda$ is symmetric
due to (\ref{eq:basic}), its contraction with (\ref{eq:1-se2}) provides
\begin{equation}
  \label{eq:c1-se2}
  3\;\xi_\lambda\nabla^\rho \nabla^\sigma \xi^\lambda
\nabla_\rho\left( \xi_{[\gamma}\xi_{\sigma;\mu]}\right)=
-2\;\xi_\lambda\nabla^\rho \nabla^\sigma \xi^\lambda
\nabla_\rho \nabla_\sigma \xi_{[\mu}\xi_{\gamma]}.
\end{equation}
Let $M^{\{\Omega\}\rho\mu\sigma}$ be any tensor with an arbitrary
number of indices (denoted by $\{\Omega\}$) plus three indices such that
$M^{\{\Omega\}[\rho\mu\sigma]}=M^{\{\Omega\}\rho(\mu\sigma)}=0$.
After relabelling the indices, the contraction of (\ref{eq:1-se2})
with $M^{\{\Omega\}\rho\mu\sigma}$ and $\xi^\gamma$ gives
\begin{equation}
  \label{eq:3-se8}
 3\;\xi^\beta M^{\{\Omega\}\sigma\rho\tau}
\nabla_\sigma\left( \xi_{[\beta}\xi_{\tau;\rho]}\right)
=
\xi^\lambda \xi_\lambda \nabla_\sigma \nabla_\rho \nabla_\tau
M^{\{\Omega\}\sigma\rho\tau} +
2\;\xi^\beta\nabla_\sigma \nabla_\tau \xi_\beta \xi_\rho
M^{\{\Omega\}\sigma\rho\tau}.
\end{equation}
Replacing now
$M^{\{\Omega\}\sigma\rho\tau}$ by $R^{\rho\tau\sigma}{}_{[\mu}\xi_{\gamma]}$
and using (\ref{eq:basic}) and (\ref{eq:c1-se2}) we find
\begin{eqnarray}
  \label{eq:3-se8+c1-se2}
 3\;\xi^\beta R^{\rho\tau\sigma}{}_{[\mu}\xi_{\gamma]}
\nabla_\sigma\left( \xi_{[\beta}\xi_{\tau;\rho]}\right)
=
\xi^\lambda \xi_\lambda \nabla_\sigma \nabla_\rho \xi_\tau
R^{\rho\tau\sigma}{}_{[\mu}\xi_{\gamma]}-
3\;\xi_\lambda\nabla^\rho \nabla^\sigma \xi^\lambda
\nabla_\rho\left( \xi_{[\gamma}\xi_{\sigma;\mu]}\right).
\end{eqnarray}
The final step is just to compute the linear combination
(\ref{eq:jotaxi})$+$(\ref{eq:c1-se2})$-\frac{1}{2}$
(\ref{eq:3-se8+c1-se2}) to obtain
$$
\xi^\alpha \xi^\beta \xi^\lambda B_{\alpha\beta\lambda[\mu}\xi_{\gamma]}=
\frac{3}{2}\left\{ -\xi_\lambda\nabla^\rho \nabla^\sigma \xi^\lambda
\nabla_\rho\left( \xi_{[\gamma}\xi_{\sigma;\mu]}\right)+
\xi^\beta R^{\rho\tau\sigma}{}_{[\mu}\xi_{\gamma]}
\nabla_\sigma\left( \xi_{[\beta}\xi_{\tau;\rho]}\right)
\right\},
$$
which, after using (\ref{eq:basic}) and some index manipulation,
gives the desired identity  (\ref{eq:xixixixi}).\fin

Note that $\xi^\alpha \xi^\beta \xi^\lambda B_{\alpha\beta\lambda\mu}=
\xi^\alpha \xi^\beta \xi^\lambda B_{(\alpha\beta\lambda)\mu}=
\xi^\alpha \xi^\beta \xi^\lambda B_{(\alpha\beta\lambda\mu)}$
due to the symmetries in (\ref{belsym}). If $\vec\xi$
is hypersurface-orthogonal, i.e. $\xi_{[\rho} \xi_{\beta;\mu]}=0$,
Result \ref{teo:belcurrents1}
explicitly follows at once as a corollary of formula (\ref{eq:xixixixi}).


\begin{identity}
Let $\xiv$ and $\etav$ be two Killing vector fields in $\espaitemps$
generating a two dimensional group of isometries acting
on non-null surfaces. Taking into account
(\ref{eq:w}), (\ref{eq:det}),
(\ref{eq:Sigma})-(\ref{eq:tOmega}) in the Appendix, the four associated currents
can be expressed as follows
\begin{eqnarray}
&&  -3 \det^2\,j_{[\alpha}\left(\xiv_A,\xiv_B\right) w_{\mu\nu]}=\nonumber\\
&&2\frac{3!}{\det}\left\{
\left(\xi_{A}^{(\lambda}
\xi_A{}_\gamma w^{\tau)\gamma} \xi_B^\epsilon
+2\xi_{B}^{(\lambda}
\xi_A{}_\gamma w^{\tau)\gamma} \xi_A^\epsilon\right)
(\eta_\lambda\tilde\Omega_{1}{}_\rho-\xi_\lambda\tilde\Omega_{2}{}_\rho)
R_\tau{}^\rho{}_\epsilon{}^\sigma
\teta_{\sigma[\alpha}w_{\mu\nu]}\right.+\nonumber\\
&&\left.+
\left(\xi_{A}^{(\lambda}
\xi_A{}_\gamma w^{\tau)\gamma} \xi_B{}_\beta w^{\epsilon\beta}
+2\xi_{A}^{(\lambda}
\xi_B{}_\gamma w^{\tau)\gamma} \xi_A{}_\beta w^{\epsilon\beta}\right)
(\eta_\lambda\tilde\Sigma_{1}{}_\rho-\xi_\lambda\tilde\Sigma_{2}{}_\rho)
R_\tau{}^\rho{}_\epsilon{}_{[\alpha}w_{\mu\nu]}\right\}+\nonumber\\
&&+\frac{3}{2}\left[
\left(2\, (\xiv_A\cdot\xiv_A)\tilde\Sigma_B{}_\tau+
4 (\xiv_A\cdot\xiv_B)\tilde\Sigma_A{}_\tau\right) w^{\lambda\sigma}
\right.+\nonumber\\
&&\left.+\left((\xiv_A\cdot\xiv_A)\tilde\Omega_B{}_\tau+
2 (\xiv_A\cdot\xiv_B)\tilde\Omega_A{}_\tau\right)\teta^{\lambda\sigma}
\right]R_{\lambda\sigma}{}^\tau{}_{[\alpha} w_{\mu\nu]}+\nonumber\\
&&-\frac{1}{8}\left(2 w^{\lambda\sigma} + \teta^{\lambda\sigma}\right)
w^{\beta\gamma}R_{\beta\gamma\lambda\sigma}
\left((\xiv_A\cdot\xiv_A) \xi_B^\tau+2(\xiv_A\cdot\xiv_B) \xi_A^\tau\right)
(\eta_\tau\Sigma_{1}{}_{[\alpha} w_{\mu\nu]}
-\xi_\tau\Sigma_{2}{}_{[\alpha} w_{\mu\nu]}),\hspace{1cm}
\label{eq:collons}
\end{eqnarray}
where the indices $A,B,C$ take the values $1,2$
to denote $\xiv$ and $\etav$ respectively, and $*$ denotes the usual
Hodge dual.
\end{identity}
\proof Not to overwhelm the text, we prefer to present the proof
in an Appendix.\fin

Clearly, if the $G_2$ isometry group is Abelian and acts orthogonally
transitively, then $\bm\Sigma_A\wedge \bm w=\bm\Omega_A\wedge \bm w=0$
and $\tilde{\bm\Sigma}_A=\tilde{\bm\Omega}_A=0$ for $A=1,2$,
as follows from (\ref{eq:Sigma})-(\ref{eq:tOmega}), and Result
\ref{teo:belcurrents2} is recovered.


Let us introduce the notation $\vec{j}_\Upsilon$, ($\Upsilon=1,2,3,4$),
for the four Bel
currents appearing in (\ref{4}). Then, Result \ref{teo:belcurrents2} implies
that
\be
\vec{j}_\Upsilon=a_{\Upsilon}(x)\vec{\xi}+
b_{\Upsilon}(x)\vec{\eta}.
\label{eq:prop}
\ee
The functions $a_\Upsilon$ and $b_\Upsilon$ are not arbitrary.
Note that, given that
$\vec\xi$ and $\vec\eta$ are Killing vectors, then
$\lie_{\vec\xi}R\, {}^{\alpha}{}_{\beta\lambda\mu}=
\lie_{\vec\eta}R\, {}^{\alpha}{}_{\beta\lambda\mu}=0$, and, therefore,
\be
\lie_{\vec\xi}B\, {}_{\alpha\beta\lambda\mu}=
\lie_{\vec\eta}B\, {}_{\alpha\beta\lambda\mu}=0.
\label{lieBel}
\ee
Hence, if we compute the Lie derivative with respect to both Killing vector
fields of equation (\ref{eq:prop}), by using (\ref{lieBel}) and the definition
(\ref{eq:current}) of Bel currents we deduce
$$
0=\lie_{\vec\xi}\vec{j}_\Upsilon=(\lie_{\vec\xi}\ a_{\Upsilon})\ \vec\xi +
(\lie_{\vec\xi}\ b_{\Upsilon})\ \vec\eta
$$
and a similar relation replacing $\lie_{\vec\xi}$ by $\lie_{\vec\eta}$.
In consequence, the functions $a_\Upsilon$ and $b_\Upsilon$ are restricted by
$$
\lie_{\vec\xi}\ a_{\Upsilon}=\lie_{\vec\xi}\ b_{\Upsilon}=
\lie_{\vec\eta}\ a_{\Upsilon}=\lie_{\vec\eta}\ b_{\Upsilon}=0.
$$

Taking now the divergence of equation (\ref{eq:prop}) and using this result
we arrive at
$$
\nabla_\rho j^{\rho}_\Upsilon=0\,\,\,\,\,\,\,\, \Upsilon=1,2,3,4\ .
$$
The case with only one hypersurface-orthogonal Killing vector can be treated
as a special case of equation (\ref{eq:prop}) with $\Upsilon=1$ and
$b_{1}=0$. All these results can be summarised as follows:
\begin{theorem}
\begin{enumerate}
\item If $\vec \xi$ is a hypersurface-orthogonal
 Killing vector then its corresponding Bel
current $\vec\CBel{\xi}{}{}{}{}$ is proportional
to $\vec\xi$ and divergence-free.
\item If $\vec \xi$ and $\vec \eta$ are two commuting Killing vectors
acting orthogonally transitively
on non-null surfaces, then their four corresponding Bel currents
$\vec{j}_\Upsilon$ lie in the 2-planes generated by $\{\vec\xi ,\vec\eta\}$ and
are divergence-free.
\end{enumerate}
\label{key}
\end{theorem}
These results are very interesting, and reinforce, once more, the various
analogies between the energy-momentum tensors and some of the
superenergy tensors. Nevertheless, as remarked in the Introduction,
they forbid the existence of {\em mixed} superenergy currents if they
are to be constructed with Killing fields of the type appearing in the
results.
Unfortunately, most of the known Einstein-Klein-Gordon
exact solutions contain such types of Killing vector fields. In order
to get a flavour of the possible interpretation of the s-e currents,
we are going to present some simple cases of pure divergence-free currents in
the next section \ref{R-N}, leaving the more difficult mixed case for
section \ref{mixed}, in which an explicit solution of the field
equations without the above mentioned properties for the Killing vector fields
will be used.

\section{Bel and Bel-Robinson currents in Reissner-Nordstr\"om and 
Schwarzschild spacetimes}
\label{R-N}
Let us consider the Reissner-Nordstr\"om spacetime, whose line-element, in
standard spherical coordinates $\{t,r,\theta,\varphi\}$, reads 
\cite{HE,MTW,KRAM}
\be
ds^2=-\left(1-\frac{2m}{r}+\frac{q^2}{r^2}\right)dt^2+
\left(1-\frac{2m}{r}+\frac{q^2}{r^2}\right)^{-1}dr^2+r^2d\Omega^2
\label{RN},
\ee
where $m$ and $q$ are arbitrary constants representing the total
mass and electric
charge of the particle creating the gravitational field, and
$d\Omega^2=d\theta^2+\sin^2\theta\ d \varphi^2$ is the
canonical line-element in the unit 2-sphere. We only consider the exterior
asymptotically flat region defined by $1-\frac{2m}{r}+\frac{q^2}{r^2}>0$ which
restricts the range of the coordinate $r$ to $r>0$ if $m^2<q^2$, or to
$r>r_+\equiv m+\sqrt{m^2-q^2}$ if $m^2\geq q^2$. The metric (\ref{RN}) is the
general spherically symmetric solution of the Einstein-Maxwell equations.
The electromagnetic field is given by $\bm{F}=(q/r^2)\ dt\wedge dr$, and
its energy-momentum tensor expressed (in natural units $8\pi G=c=1$
where $c$ is the speed of light in vacuum and $G$ is the gravitational constant)
in the coordinate basis is
\begin{equation}
T^{\mu}{}_{\nu}=\frac{q^2}{r^4}\times\mbox{diag}\left\{-1,-1,1,1\right\}.
\label{RNem}
\end{equation}
The particular case with $q=0$ is the vacuum Schwarzschild
solution (then $r>2m$).

In the chosen range of coordinates, the metric is static
($\xiv=\partial/\partial t$
is a hypersurface-orthogonal timelike Killing vector)
and spherically symmetric.
We enumerate the four Killing vectors $\xiv_L$, ($L=1,2,3,4$) as
\be
\xiv_1=\frac{\partial}{\partial t},~~
\vec\xi_2 = \sin \varphi\frac{\partial}{\partial \theta}+
\cos \varphi \cot \theta \frac{\partial}{\partial \varphi},~~
\vec\xi_3 = \cos \varphi\frac{\partial}{\partial \theta}-
\sin \varphi \cot \theta \frac{\partial}{\partial \varphi},~~
\vec\xi_4
=\frac{\partial}{\partial \varphi}.
\label{RNkillings}
\ee

We want to study the Bel (and Bel-Robinson for the $q=0$-case) currents
constructed with the Killing vectors (\ref{RNkillings}) for
these spacetimes, together with the currents arising from the energy-momentum
tensor. These will lead to some conserved quantities. Of course, one expects
that all the conserved quantities will be just functions of the constants
$m$ and $q$, but our aim is to ascertain which particular combinations of $m$
and $q$ arise at the energy and superenergy levels. To find them
we compute explicitly all the Bel currents:
\begin{eqnarray}
&&\bm{j}(\xiv_1)= - \frac{1}{r^6}\left(1-\frac{2m}{r}+\frac{q^2}{r^2}\right)
\left[6\left(m-\frac{q^2}{r}\right)^2+\frac{q^4}{r^2}\right]\bm\xi_1,
\label{j1}\\
&&\CBef{\xi_a}{\xi_1}{}=
\frac{1}{3 r^6}\left(1-\frac{2m}{r}+\frac{q^2}{r^2}\right)
\left[12\left(m-\frac{q^2}{r}\right)^2+\frac{q^4}{r^2}\right]\bm\xi_a,\nonumber\\
&&\CBf{\xi_a}{\xi_b}{\xi_1}{}=
- \frac{1}{3 r^6}\left(\vec\xi_a\cdot \vec\xi_b\right)
\left[12\left(m-\frac{q^2}{r}\right)^2+\frac{q^4}{r^2}\right]\bm\xi_1,
\label{j2}\\
&&\CBf{\xi_a}{\xi_b}{\xi_c}{}=
\frac{1}{3 r^6}
\left[6\left(m-\frac{q^2}{r}\right)^2+\frac{q^4}{r^2}\right]
\left\{\left(\vec\xi_a\cdot \vec\xi_b\right)\bm\xi_c+
\left(\vec\xi_b\cdot \vec\xi_c\right)\bm\xi_a+
\left(\vec\xi_c\cdot \vec\xi_a\right)\bm\xi_b
\right\},\nonumber
\end{eqnarray}
where $a,b,c\in\{2,3,4\}$.
Similarly, the energy-momentum currents defined in (\ref{eq:ct}) can be
computed on this metric and they read:
\begin{eqnarray}
&&\CTf{\xi_1}=- \frac{q^2}{r^4}\bm\xi_1\label{jj1}\\
&&\CTf{\xi_a}= \frac{q^2}{r^4}\bm\xi_a.\nonumber
\end{eqnarray}

It can be easily checked that all the above currents are divergence-free, and
therefore they lead to a set of conserved quantities via Gauss' theorem applied
to appropriate compact 4-volumes of the spacetime. For illustration purposes,
we are going to compute now these conserved quantities for the simple but
relevant case in which the compact region is taken to be bounded by two
$t=$const.\ hypersurfaces, say $t=t_1$ and $t=t_2$, $t_2>t_1$, and
two $r=\hbox{const}$.\
hypersurfaces, say $r=r_1$ and $r=r_2$, $r_2>r_1$. We call this region
$\K$. Applying Gauss' theorem we immediately get
$$
\int_{\partial \K}A^{\mu}n_{\mu}\, \d^3\sigma =0,
$$
where  $\vec A$ represents any of the $\vec j$ or $\vec{\cal J}$ computed
above, $\partial \K$ denotes the boundary of $\K$, $\bm{n}$ is the outward
unit normal to $\K$ and $\d^3\sigma$ is the canonical volume
3-element
on $\partial \K$. Clearly,
$\partial \K =\{t=t_1,r_1<r<r_2\}\cup \{t=t_2,r_1<r<r_2\}
\cup \{r=r_1,t_1<t<t_2\}\cup \{r=r_2,t_1<t<t_2\}$,
so that the corresponding
unit normals are proportional to $dt$ for the first two regions, and to $dr$ for
the remaining two. From the explicit expressions of the currents, we know that
none of them has a non-zero component along $dr$, and therefore
the integrals on
$\{r=r_1\}\cup \{r=r_2\}$ vanish. Thus, only the integrals on $\{t=t_1\}$ and
$\{t=t_2\}$ remain, and thus they must be equal. In other words, the
integrals
\be
\int_{t=\mbox{const.}}A^{\mu}n_{\mu}\, \d^3\sigma \label{otra},
\ee
taken over any portion of a $t=$const.\ hypersurface bounded by the values
$r_1$ and $r_2$, are constant in the sense that they are independent of the
particular $t=$const.\ hypersurface. It is quite remarkable that the 
integrand in (\ref{otra}) for the case of the electromagnetic current
(\ref{jj1}) is {\em not} the standard expression for the electromagnetic
energy density, taken usually as $T^{\mu\nu}n_{\mu}n_{\nu}$. However,
(\ref{otra}) leads to simpler results and to expressions which are apparently 
correct (at least in a naive comparison with the analogous ones in the
classical theory) such as (\ref{otra2}) and (\ref{otra3}) found later. 
Thus, one wonders if, in a stationary frame of reference defined by $\xiv$, 
the correct expression for the electromagnetic energy density taken over an 
extended region of the spacetime ---and not only at a point--- should be 
in fact $T^{\mu\nu}\xi_{\mu}n_{\nu}$.

Coming back to (\ref{otra}), and by noting that 
$\bm{n}=-\left(1-\frac{2m}{r}+\frac{q^2}{r^2}\right)^{1/2}dt$
and that $\d^3\sigma =\left(1-\frac{2m}{r}+\frac{q^2}{r^2}\right)^{-1/2}
r^2\sin\theta drd\theta d\varphi$, those integrals reduce to
\begin{eqnarray}
\int_0^{2\,\pi}d{\varphi}\int_0^{\pi}\sin\theta\, d{\theta}\int_{r_1}^{r_2}
(-A^{t}) r^2 dr,
\end{eqnarray}
where $A^t$ denotes the $t$-component of $\vec{A}$.
Observe that, once again, among the explicit expressions for $\vec j$ and
$\vec{\cal J}$, only those involving $\vec\xi_1$ have a non-zero $t$-component,
and therefore they are the only ones which may give {\it non-trivial}
conserved quantities: all other currents provide constants which are
simply identically zero. Actually, not even all of the currents involving
$\vec\xi_1$ provide a non-zero constant, because the angular integrals can also
vanish. In summary, it is easy to check that
the only non-trivial constants arise
from the currents (\ref{jj1}), (\ref{j1}) and the three
with $a=b$ in (\ref{j2}),
these last three being, in fact, equal.
Denoting them by ${\cal Q}_1$,
$Q_1$ and $Q_2$ respectively, they clearly depend  on the values
of $r_1$ and $r_2$. Their explicit expressions are:
\begin{eqnarray}
{\cal Q}_{1}=4\,\pi \,{q^2}\left(\frac{1}{r_1}-\frac{1}{r_2}\right),
\hspace{4cm}\label{otra2}\\
Q_{1}=\left.\frac{4\,\pi}{15\,r^7}\left(45m^3r^3-90m^2q^2r^2-30m^2r^4+65mq^4r+
45mq^2r^3-21q^4r^2-15q^6\right)\right\vert_{r_1}^{r_2}\, ,\nonumber\\
Q_2=-\left.\frac{8\pi}{27r^3}\left(36m^2r^2-36mq^2r+13q^4\right)
\right\vert_{r_1}^{r_2}\, .\hspace{2cm}\nonumber
\end{eqnarray}
It is also remarkable that these three expressions are strictly 
positive. This is a general result---independent of the spacetime if 
it is stationary--- for the currents of type ${\cal Q}_{1}$ and 
$Q_{1}$, and follows from the dominant energy condition satisfied by 
$T^{\mu\nu}$ in the first case, and from the analogous dominant property 
satisfied by the Bel tensor \cite{PR,Ber2,Ber3,supertotxo} in the 
second, as the corresponding integrands are positive. This was noted 
for the Bel-Robinson case for instance in \cite{CK} (see also the 
recent generalisation in \cite{J}). Nevertheless, 
there is no similar reasoning for $Q_{2}$ and its positivity can only 
be inferred from the explicit expression of the local current (\ref{j2}).

If we want to obtain constants associated to the spacetime, we can also take
the limit cases in which the above expressions are computed over a whole
slice $t=$const.\ by taking  $r_2\rightarrow \infty$ and the minimum possible
value for $r_1$. There appear then two different situations depending on
whether $m^2\geq q^2$ or not. If $m^2<q^2$, then the minimum value for $r_1$ is
$r_1=0$, and clearly all the above constants diverge. This is natural and
analogous to what happens in flat spacetime if we integrate the energy over
the whole space. However, for the case with $m^2\geq q^2$, the existence of the
event horizon at $r_+=m+\sqrt{m^2-q^2}$ provides a finite minimum value for
$r_1$, and this leads to finite conserved constants. A simple computation
leads to
\begin{eqnarray*}
{\cal Q}_{1}&=&\frac{4\,\pi \,{q^2}}{r_+}, \\
Q_{1}&=&\frac{4\,\pi}{15\,r_+}\left( 6 -14\frac{m}{r_+}+14\frac{m^2}{r_+^2}
-5\frac{m^3}{r_+^3}\right),
\\
Q_{2}&=&\frac{8\,\pi}{27}r_+
\left( 13-16\frac{m}{r_+}+16\frac{m^2}{r_+^2}\right).
\end{eqnarray*}
For the particular case of the Schwarzschild spacetime ($q=0$), they become
simply
{\setlength\arraycolsep{0.1pt}
$$
{\cal Q}_{1}=0, \hspace{1cm}
Q_1=\frac{\pi}{4m}, \hspace{1cm} Q_2=\frac{16\,\pi\,m}{3}\, .
$$
An obvious question arises: are these {\em particular} combinations of $m$ and
$q$ special in any sense? And if yes, why?
Clearly, the meaning of ${\cal Q}_1$ is the total electrostatic energy
of the Reissner-Nordstr\"om black hole with respect to the static observer.
However, in order to check this simple statement one has to put back all the
physical constants which were implicitly taken to be unity.
Denoting by $M$ and $Q$
the total mass and charge in the correct units, respectively, we have
$m=GM/c^2$ and $q^2=GQ^2/c^4$.
Now, the energy-momentum tensor (\ref{RNem}) must
be multiplied by $c^4/(8\pi G)$, and therefore we finally get for the physical
${\cal Q}_1$:
\be
{\cal Q}_{1}=\frac{1}{2}\, \frac{Q^2}{r_+}.\label{otra3}
\ee
This is a satisfactory result which may provide a physical 
interpretation for ${\cal Q}_1$, as this formula is reminiscent of 
that for the energy of the electric field in classical physics.

Now, the same type of reasoning is needed for the superenergy quantities
$Q_1$ and $Q_2$. To that end, we need to know which type of physical 
dimensions are carried by the superenergy tensors. This can be deduced 
from several independent
works \cite{HS,Berg,Tey,Kri}, and as is explained in
\cite{supertotxo} (p.\ 2820)
the correct physical units for the {\em physical} Bel tensor seem to 
be energy density
per unit surface, i.e. $ML^{-3}T^{-2}$. This means that we have to multiply the
Bel tensor by $\kappa c^4/G$ to get the correct physical quantities, where
$\kappa$ is a pure number to be chosen. Doing that we can
obtain the following expression for the physical
$Q_1$:
\begin{equation}
Q_1=\kappa\frac{4\pi }
{r_+^2}\left(\frac{1}{4}\, Mc^2+\beta \frac{Q^2}{r_+}\right)=
\kappa \frac{4\pi }{r_+^2}
\left(\frac{1}{4}\, Mc^2+2 \beta {\cal Q}_1\right)
\label{Q1}
\end{equation}
where $\displaystyle{\beta =
-\frac{1}{2}\left(\frac{1}{3}\frac{m^2}{r_+^2}-
\frac{23}{30}\frac{m}{r_+}+\frac{4}{5}\right)}$
is a dimensionless quantity.
Of course, since (\ref{Q1}) is written
in terms of three non-independent quantities, $M$, $Q$ and $r_+$,
it is not unique. It only provides a possible good choice.
For the particular case of
Schwarzschild's solution ($q=0 \Rightarrow r_+=2m$) the previous quantity
reduces simply
to
$$
Q_1=\kappa\frac{\pi}{4m^2} M c^2 \, .
$$

We note in passing that if we chose $\kappa =(2\pi)^{-2}$
we could rewrite this as
\begin{equation}
Q_1=\frac{M c^2}{{\cal A}}
\label{Q1'}
\end{equation}
where ${\cal A}$ is the area of the horizon.
Nevertheless, such an amusing or inspiring interpretation for $Q_1$ is
not possible in the general case (\ref{Q1}) due to the factor $\beta$.

A possible way to try and obtain alternative, more complete, 
expressions of type (\ref{Q1}) is by using the existence of independent
superenergy of pure electromagnetic origin, see \cite{supertotxo}. In 
order to check this possibility, we can use any of the s-e tensors for the 
electromagnetic field (the most general one depends on six arbitrary 
constants, see \cite{S,supertotxo}) because the corresponding current 
constructed with $\xiv_{1}$ is always the same, as can be easily 
checked. Thus, we can use
the basic s-e tensor $T_{\alpha\beta\lambda\mu}\{\nabla_{[1]}F_{[2]}\}$
for $\nabla F$, given by \cite{S,supertotxo}
\bea
E_{\alpha\beta\lambda\mu}\equiv
\nabla_{\alpha}F_{\lambda\rho}\nabla_{\beta}F_{\mu}{}^{\rho}+
\nabla_{\alpha}F_{\mu\rho}\nabla_{\beta}F_{\lambda}{}^{\rho}-
g_{\alpha\beta}\nabla_{\sigma}F_{\lambda\rho}
\nabla^{\sigma}F_{\mu}{}^{\rho}- \nonumber \\
-\frac{1}{2}g_{\lambda\mu}
\nabla_{\alpha}F_{\sigma\rho}\nabla_{\beta}F^{\sigma\rho}+
\frac{1}{4}g_{\alpha\beta}g_{\lambda\mu}
\nabla_{\tau}F_{\sigma\rho}\nabla^{\tau}F^{\sigma\rho} \, .
\label{seF}
\eea
Its symmetry properties are
\bean
E_{\alpha\beta\lambda\mu}=E_{(\alpha\beta)(\lambda\mu)}
\eean
but it is not symmetric in the exchange of $\alpha\beta$ with 
$\lambda\mu$ (a tensor with that symmetry was found many years ago by 
Chevreton \cite{Ch,Tey}; Chevreton's tensor is simply proportional to
$E_{\alpha\beta\lambda\mu}+E_{\lambda\mu\alpha\beta}$. In what 
follows, one can use either of these tensors or, for that matter, its 
completely symmetric part). The current associated to (\ref{seF}), and 
analogous to (\ref{j1}), can be easily computed to produce
\be
E^{\alpha}{}_{\beta\lambda\mu}\xi^{\beta}_{1}\xi^{\lambda}_{1}\xi^{\mu}_{1}=
-\frac{3q^2}{r^6}\left(1-\frac{2m}{r}+\frac{q^2}{r^2}\right)^2\, 
\xi^{\alpha}_{1} \label{j1F},
\ee
so that it is clearly divergence-free too. Thus, by a similar procedure as above
it gives rise to a  conserved quantity  when integrated 
over any $t=$constant spatial section. Its value is
\bea
Q_{1}^{(F)}=\frac{4\pi\, q^2}{35\, r_{+}^3}
\left(4\frac{m^2}{r_{+}^2}-11\frac{m}{r_{+}}+8\right) =
\frac{4\pi}{35\, r_{+}}\left(8\frac{m^3}{r_{+}^3}-26\frac{m^2}{r_{+}^2}+
27\frac{m}{r_{+}}-8 \right). \label{Q1F}
\eea
Therefore, by forming linear combinations with positive coefficients 
$c_{1}$ and $c_{2}$ of the respective s-e tensors we can get a 
conserved quantity given by $c_{1}Q_{1}+c_{2}Q_{1}^{(F)}$ which, when 
the physical units have been restored, becomes
\begin{eqnarray*}
c_{1}Q_{1}+c_{2}Q_{1}^{(F)}=\kappa\frac{4\pi }
{r_+^2}\left(A\, Mc^2+B \frac{Q^2}{r_+}\right) 
\end{eqnarray*}
with
\begin{eqnarray*}
A=a+(b-2a)\frac{m}{r_{+}}+(c_{1}-2b)\frac{m^2}{r_{+}^2},\hspace{3cm}\\
B=\frac{8}{35}c_{2}-\frac{2}{5}c_{1}+(a+\frac{2}{15}c_{1}-
\frac{11}{35}c_{2})\frac{m}{r_{+}}+
(b-\frac{2}{3}c_{1}+\frac{4}{35}c_{2})\frac{m^2}{r_{+}^2}.
\end{eqnarray*}
Here $a$ and $b$ are spureous constants (due to the fact that 
$1-2m/r_{+}+q^2/r_{+}^2 =0$), so that these expressions can be 
simplified on using this freedom. One can try to restrict the values of the 
constants by using other kind of arguments. For instance, there are independent 
results that prove the conservation of the simply
added superenergy along shock-wave discontinuities for 
Einstein-Maxwell spacetimes \cite{Lic} (see also subsection 7.3 in 
\cite{supertotxo} and references therein), hence
it seems reasonable to put in any case 
$c_{1}=c_{2}$. Setting then $c_{1}=c_{2}=1$ and, as an example,
choosing $a$ and $b$ such that $A$ does not depend on $m/r_+$ 
so that the case $Q=0$ is explicitly recovered, we get
$$
Q_{1}+Q_{1}^{(F)}=\kappa\frac{4\pi }
{r_+^2}\left(\frac{1}{4} M c^2+B\frac{Q^2}{r_+}
\right),
$$
where now $\displaystyle B=-\frac{1}{35}\left(
\frac{11}{6}\frac{m^2}{r_+^2}-\frac{29}{12}\frac{m}{r_+} +6\right)$.
Other choices for $a$ and $b$ lead to similar formulas.
Thus, the interpretation of $Q_{1}$ and, in general, 
of its combinations with $Q_{1}^{(F)}$, remains somewhat obscure. 

For the sake of completeness, we include now the physical $Q_2$ 
(note that the corresponding $Q_{2}^{(F)}$ 
vanishes, so that in this case the linear combinations of the 
superenergy tensors always lead to a quantity proportional to $Q_{2}$):
$$
Q_2=8\pi\kappa \left(\delta M c^2 + \gamma  \frac{Q^2}{r_+}\right)=
8\pi\kappa\left(\delta M c^2 + 2 \gamma  {\cal Q}_1\right),
$$
where now $\delta=2(8/27-k)m/r_+ +k  +10/27$ and
$\displaystyle\gamma=k\, m/r_+ -13/27$ and $k$ is a disposable 
constant similar to the previous $a$ and $b$.
Choosing again $\delta$ not to depend on $m/r_+$, one gets
$$
Q_2=\kappa
\frac{16\pi}{3}\left[ M c^2 + 
\frac{1}{9} \left(8\, \frac{m}{r_+}-13\right)
{\cal Q}_1\right].
$$
The units of this quantity are those of energy, that is 
$ML^2T^{-2}$, but given its origin and construction one has doubts 
about whether considering it as such or as a kind of ``supermomentum''
per unit surface: $(M L^2 T^{-2}) L^2 /L^2$. In the Schwarzschild case,
$q=0$, one gets then
$$
Q_2=\kappa\frac{16\pi}{3}M c^2.
$$ 

\section{Conserved mixed currents}\label{mixed}
As discussed in the Introduction, and remarked at the end of the previous
section, if there are non-gravitational fields present in the spacetime 
then they may also contribute to the total superenergy tensor. 
One important question is whether or
not the total s-e current thus constructed may be divergence-free. Of course,
this depends on the field equations for the particular matter fields involved.
In \cite{S2,supertotxo}, it was proved in full generality that these 
divergence-free {\em mixed} s-e currents can always be constructed 
for the simple case of a
minimally coupled scalar field (whenever there is a Killing vector in the
spacetime). We now summarize here the main results in \cite{S2,supertotxo}
concerning the massless case, which is the one relevant for
our present purposes.

Let $\phi$ be a massless scalar field. The s-e tensor for the scalar field is
the basic s-e tensor $T_{\alpha\beta\lambda\mu}\{\nabla_{[1]}\nabla_{[1]}\phi\}$ 
for $\nabla\nabla\phi$ and can be easily constructed following the general 
definition of \cite{supertotxo}. The result is (see \cite{S,supertotxo,S2,Tey}):
\begin{eqnarray}
S_{\alpha\beta\lambda\mu}\equiv \nabla_{\alpha}\nabla_{\lambda}\phi
\nabla_{\mu}\nabla_{\beta}\phi +\nabla_{\alpha}\nabla_{\mu}\phi
\nabla_{\beta}\nabla_{\lambda}\phi -\nonumber \hspace{2cm}\\
-g_{\alpha\beta}\nabla_{\lambda}\nabla^{\rho}\phi
\nabla_{\mu}\nabla_{\rho}\phi -g_{\lambda\mu}\nabla_{\alpha}\nabla^{\rho}\phi
\nabla_{\beta}\nabla_{\rho}\phi +\frac{1}{2}g_{\alpha\beta}g_{\lambda\mu}
\nabla_{\sigma}\nabla_{\rho}\phi \nabla^{\sigma}\nabla^{\rho}\phi
\label{s-eS}
\end{eqnarray}
from where we immediately deduce
$$
S_{\alpha\beta\lambda\mu}=S_{(\alpha\beta)(\lambda\mu)}=
S_{\lambda\mu\alpha\beta}.
$$
One can straightforwardly compute the divergence of the scalar field s-e tensor
$$
\nabla_{\alpha}S^{\alpha}{}_{\beta\lambda\mu}=
2\nabla_{\beta}\nabla_{(\lambda}\phi
R_{\mu)\rho}\nabla^{\rho}\phi-
g_{\lambda\mu}R^{\sigma\rho}\nabla_{\beta}\nabla_{\rho}\phi\nabla_{\sigma}\phi-
\nabla_{\sigma}\phi (2\nabla^{\rho}\nabla_{(\lambda}\phi
R^{\sigma}{}_{\mu)\rho\beta}+g_{\lambda\mu}R^{\sigma}{}_{\rho\beta\tau}
\nabla^{\rho}\nabla^{\tau}\phi )
$$
so that we realize that the s-e tensor (\ref{s-eS}) is divergence-free in
flat spacetime, that is to say, in the absence of gravatational field. This
allows to construct conserved currents for the scalar field in flat spacetime.
They are built in a similar way to those formed with the energy-momentum
tensor or the Bel currents (\ref{eq:current}). Following \cite{S2,supertotxo,Tey},
one can define the
{\em scalar-field s-e current} with respect to any three Killing vector fields
$\vec \xi^{}_1$, $\vec \xi^{}_2$, $\vec \xi^{}_3$ as
$$
\gimel_{\mu}\left(\vec{\xi}^{}_1,\vec{\xi}^{}_2,\vec{\xi}^{}_3\right)\equiv
S_{(\alpha\beta\lambda)\mu}\,\xi^{\alpha}_1\xi^{\beta}_2\xi^{\lambda}_3=
S_{(\alpha\beta\lambda\mu)}\,\xi^{\alpha}_1\xi^{\beta}_2\xi^{\lambda}_3.
$$
Nevertheless, these currents are not divergence-free in curved spacetimes.
In other words, they are not divergence-free if one takes into account the
gravitational field created by the scalar field. This was to be 
expected: if the s-e concept is to have any physical meaning at all, then the
{\em total} s-e currents, involving both the gravitational and the scalar fields,
are the ones to be conserved. And this was actually proved in
\cite{S2,supertotxo}:
if the Einstein-Klein-Gordon field equations are satisfied, that is
$$
R_{\mu\nu}=\nabla_{\mu}\phi\nabla_{\nu}\phi \hspace{1cm} \Longrightarrow
\hspace{1cm} \nabla_{\mu}\nabla^{\mu}\phi =0
$$
then the sum of the Bel current and the scalar-field s-e current, which will be
written as
$$
\vec J(\xiv_1,\xiv_2,\xiv_3)\equiv \vec j(\xiv_1,\xiv_2,\xiv_3)+
\vec{\gimel}(\xiv_1,\xiv_2,\xiv_3)
$$
is a divergence-free vector field:
$$
\nabla_{\alpha}J^{\alpha}(\xiv_1,\xiv_2,\xiv_3)=0.
$$
Observe that $\vec J$ is explicitly defined by
$$
J_{\mu}\left(\vec{\xi}^{}_1,\vec{\xi}^{}_2,\vec{\xi}^{}_3\right)\equiv
\left(B_{(\alpha\beta\lambda)\mu} + S_{(\alpha\beta\lambda)\mu}\right)
\xi^{\alpha}_1\xi^{\beta}_2\xi^{\lambda}_3 \, .
$$
These are mixed conserved currents, and they lead to the conservation of
{\em mixed} quantities, containing both gravitational and scalar fields
contributions. Moreover, note that, in general, none of the two single
currents $\vec{\gimel}$ nor $\vec j$ are divergence-free separately.

We now want to produce some explicit examples of these conserved mixed
currents. To that end, we need an explicit solution of the Einstein-Klein-Gordon
equations. There are several of these in the literature ---note that sometimes 
they are considered as ``stiff fluid'' solutions.--- The problem, however, 
is that most
of the solutions have either hypersurface-orthogonal Killing vectors, or
spherical symmetry, or a $G_2$-group acting orthogonally transitively, so that
the results of section \ref{Belcurrents} apply. This means that, for these
particular solutions, the Bel and scalar-field s-e currents are, actually,
divergence-free on their own. And thus, they do not give the {\em mixed}
conservation we are seeking. What one needs, therefore, is a solution belonging 
to class A(ii) in Wainwright's classification of $G_2$ solutions 
\cite{WW1,raultesis}, that is to say, a solution with a non-orthogonally 
transitive $G_2$ group of motions.\footnote{A spacetime with 
just one {\em non-hypersurface-orthogonal} Killing vector will also do, 
but we do not know of any such solutions for the Einstein-Klein-Gordon system.}

Remarkably and fortunately, there exists some explicitly known
solutions for spacetimes belonging to class A(ii)
and with a minimally coupled massless scalar field $\phi$ as source.
These were studied by Wils \cite{Wils} who, among other
metrics of the mentioned sort, found the one-parameter family of metrics
that we are going to use here.

In general, the line element for such solutions will take the following form:
$$
ds^2=-F_0^2dt^2+F_1^2dx^2+F_2[F_3^2dy^2+F_3^{-2}(dz+F_4 dx)^2]
$$
where $F_0$, $F_1$, $F_2$ and $F_3$ and $F_4$ are functions of $t$ and
$x$ only. Wils found the mentioned exact solutions  making
separability assumptions on the functions involved, and the
solution we are interested in has
\begin{eqnarray*}
&&F_0(t,x)=F_1(t,x)=
{\sinh^{{\scriptscriptstyle 2 + \lambda }} \frac{Lt }{{\sqrt{6}}}}\,
e^{\scriptscriptstyle\frac{\left( 3 + 4\,\lambda  \right)}{6}Lx }\, ,\\
&&F_2(t,x)={\sinh \frac{Lt }{{\sqrt{6}}}}\,
e^{\scriptscriptstyle\frac{\lambda}{3}Lx }\, ,\\
&&F_3(t,x)={\sinh^{-\frac{1 +2\, \lambda}{2} } \frac{Lt }{{\sqrt{6}}}}\,
e^{\scriptscriptstyle-\frac{\left( 3 + \lambda  \right)}{6}Lx }\, ,\\
&&F_4(t,x)={\cosh \frac{Lt }{{\sqrt{6}}}}\,
\,{\sqrt{2 - \frac{4\,{\lambda }^2}{3}}}\,
e^{\scriptscriptstyle\frac{\lambda }{3}Lx}\, ,
\end{eqnarray*}
and
$$
\phi=\sqrt{4-2\,\lambda^2}\,\log \left(\tanh \frac{Lt}{2\,{\sqrt{6}}}\right),
$$
where $L$ is a positive constant defining the scale and $\lambda$ is a
parameter subject to $\lambda^2\leq 3/2$.
The range of coordinates is constrained by the singularity at
$t=0$, and thus we take $t>0$. 

Obviously, $\xiv={\partial}/{\partial y}$ is a Killing vector that satisfies
$\xif\wedge d\xif=0$, so that it is hypersurface orthogonal. From Result 
\ref{teo:belcurrents1} it follows that the Bel current
$\vec j(\vec\xi)$ is parallel to $\vec\xi$ and hence conserved.
In turn, and because the total mixed current $\vec J(\vec\xi)$ is also
conserved, the current $\vec{\gimel}(\vec\xi)$ will be divergence-free 
too. Thus, the sought s-e interchange cannot be found here.
We include the expression for $\vec J(\vec\xi)$ for completeness:
$$
\vec J(\vec\xi)=\frac{L^4}{72}\!
\left(4(\lambda^3-\lambda+3)(2\lambda+3)\sinh^2\!\frac{Lt}{\sqrt{6}}
-\lambda^4+20\lambda^2+72\lambda+68\right)
\sinh^{\scriptscriptstyle-
6\left( 2 + \lambda  \right)}\!\frac{Lt}{\sqrt{6}}
e^{-\frac{\left( 9 + 8\lambda  \right) }{3}Lx}\,\vec\xi.
$$
However, we can also use the second Killing vector, which is
$\etav={\partial}/{\partial z}$, and one has
\[
\etaf \wedge d\etaf=\frac{L}{3}\sqrt{3-2\lambda^2}\,\,
e^{\frac{
\left( 6 + 5\,\lambda  \right)}{3}\,Lx}
{\sinh^{\scriptscriptstyle 5 + 4\,\lambda } \frac{Lt }{{\sqrt{6}}}}\,\, dt
\wedge dx \wedge dz,
\]
meaning that $\etav$ will only be hypersurface orthogonal in the
extreme cases $\lambda=\pm\sqrt{3/2}$. Nevertheless, it can be checked that 
the current $\vec j(\vec\xi,\vec\eta)$,
as well as $\vec{\gimel}(\vec\xi,\vec\eta)$, is in fact proportional to 
$\vec\xi$, so that both of them are divergence-free separately once 
more. Their sum reads
\begin{eqnarray*}
\vec J(\vec\xi,\vec\eta)=&&-\frac{L^4}{648}
\left[3\left(\lambda^4+64\lambda^3+140\lambda^2+40\lambda-44
-2(3-2\lambda^2)^2\right)\sinh ^4\frac{Lt}{\sqrt{6}}\right.\\
&&\left.-4\left(14\lambda^4-21\lambda^3-97\lambda^2-30\lambda+36\right)
\sinh ^2\frac{Lt}{\sqrt{6}}\right]
\sinh^{\scriptscriptstyle-
2\left( 5 + \lambda  \right)} \frac{Lt}{\sqrt{6}}
e^{-\left( 1 + 2\lambda  \right)Lx}\,\,\,\vec\xi \, .
\end{eqnarray*}

The only left possibilities for a true exchange of s-e properties
are to be found in the remaining conserved currents $\vec J(\vec\eta)$
and $\vec J(\vec\eta,\vec\xi)$. And this is indeed the case.
Setting $x^0=t$, $x^1=x$, $x ^2=y$ and $x^3=z$, the expressions for 
their non-zero components are the following:
{\setlength\arraycolsep{0.1pt}
\begin{eqnarray*}
J^{0}(\etav,\xiv)=&&\frac{L^4}{324}
{
\left( 3 + 2\lambda  \right)
      {\sqrt{3 - 2{\lambda }^2}}\,
{\left( 4 (1 - 2{\lambda }^2) +
       3 \left( 3 - 2{\lambda }^2 \right)
\,\sinh ^{ 2} \frac{Lt }{{\sqrt{6}}}  \right) }}
\times\\
&&
{\sinh^
       {\scriptscriptstyle-(11 + 6\lambda) } \frac{Lt\, }{{\sqrt{6}}}}
{\,e^{\scriptscriptstyle-3\left( 1 + \lambda  \right)Lx }}\\
&& \\
J^{1}(\etav,\xiv)=&&-\frac{L^4}{54\sqrt{6}}
\sqrt{3 - 2{\lambda }^2}
    \left( 2\left( 3 + 2\lambda  \right)
\left( 1 - 2{\lambda }^2 \right)  +
      3\left( 1 + \lambda  \right)\left( 3 - 2{\lambda }^2 \right)
      \sinh ^2\frac{Lt}{\sqrt{6}}\right)\times \\
&&\cosh \frac{Lt }{\sqrt{6}}
\sinh^{\scriptscriptstyle-
6\left( 2 + \lambda  \right)} \frac{Lt}{\sqrt{6}}
   \, e^{\scriptscriptstyle-3\left( 1 + \lambda  \right)Lx }
\\
&&\\
J^{3}(\etav,\xiv)=&&{\frac{L^4}{648}
 \,\left.\bigg(204 - 72\lambda  -
612{\lambda }^2 - 320{\lambda }^3 +
        93{\lambda }^4 + 64{\lambda }^5 +
        4\left( 81 + 9\lambda  -181{\lambda }^2 -
\right.\right.}\\
&&{\left.\left. -89{\lambda }^3 +
50{\lambda }^4 +
           28{\lambda }^5 \right)\sinh^2
\frac{Lt }{{\sqrt{6}}} +
      6\,\left( 3 + 2\lambda  \right) {\left( 3 - 2{\lambda }^2
\right) }^2
         {\sinh^4 \frac{Lt }{{\sqrt{6}}}} \right)  }\times
\\&&
{\sinh^
       {\scriptscriptstyle-6(2 + \lambda) } \frac{Lt }{{\sqrt{6}}}}\,{
    e^{\scriptscriptstyle-\frac{\left( 9 + 8\lambda  \right) }{3}Lx}}\\
J^{0}(\etav)=&&\frac{L^4}{54}\lambda {\sqrt{3 - 2{\lambda }^2}}
    \left( 1 -\left( 3 - 2{\lambda }^2 \right)
       \sinh^2 \frac{Lt}{{\sqrt{6}}}\right)
\sinh^{\scriptscriptstyle-9 - 2\lambda } \frac{Lt}{\sqrt{6}}\,
    e^{-\frac{\left( 3 + 7\lambda  \right)Lx }{3}}\\
J^{1}(\etav)=&&-\frac{L^4}{18{\sqrt{6}}}{{\sqrt{3 - 2{\lambda }^2}}
      \left( 2 - \left( 3 -2{\lambda }^2 \right)
         {\sinh ^2 \frac{Lt}{{\sqrt{6}}}} \right)  }
      \cosh \frac{Lt}{{\sqrt{6}}}
{\, {\sinh^{-2\,\left( 5 + \lambda  \right) } \frac{Lt}{{\sqrt{6}}}}\,}\times\\&&{
    e^{\scriptscriptstyle-\frac{\left( 3 + 7\,\lambda  \right) }{3}Lx}\,
   }\\
J^{3}(\etav)=&&\frac{L^4}{216}{{
    \left( 228 + 216\,\lambda  + 44\,{\lambda }^2 - 3\,{\lambda }^4 +
      4\,\left( 18 + 9\,\lambda  + 6\,{\lambda }^2 + 9\,{\lambda }^3 +
2\,{\lambda }^4 \right) \,
       {\sinh^{ 2} \frac{Lt }{{\sqrt{6}}}} - \right.}}
\\&&{{\left.-
      2\,{\left( 3 - 2\,{\lambda }^2 \right) }^2\,
       {\sinh^4 \frac{Lt }{{\sqrt{6}}}} \right) }
{\sinh^
{\scriptscriptstyle-2(5 +\,\lambda) } \frac{Lt }{{\sqrt{6}}}}\,
    e^{\scriptscriptstyle-\left( 1 + 2\,\lambda  \right)\,Lx }}.
\end{eqnarray*}}
It can be checked by an explicit computation that these vector fields 
are divergence-free. 
We want to stress that the corresponding pure currents $\vec j(\vec\eta)$
and $\vec{\gimel}(\vec\eta)$, or $\vec j(\vec\eta,\vec\xi)$ and
$\vec{\gimel}(\vec\eta,\vec\xi)$, are {\em not} divergence-free in 
general (unless, of course, $3-2\lambda^2=0$ in which case $\{\vec\xi,\etav\}$ 
generate an orthogonal transitive $G_2$, in agreement with our 
previous results). More importantly, we must remark
that these currents $\vec J(\vec\eta)$ and $\vec J(\vec\eta,\vec\xi)$
are {\em non-trivial} in the sense that they are {\em not}
linear combinations of the Killing vectors. In other words, they would 
be very difficult to find if we did not know about the superenergy concept.
The explicit expressions for
superenergy tensors of the gravitational and scalar fields are 
essential here.

A door obviously open by our work can be stated in the form of the 
following yet unanswered question: are there similar non-trivial {\em 
mixed} superenergy currents for general Einstein-Maxwell systems?

\section*{Acknowledgements}
We are grateful to Llu\'{\i}s Bel for many interesting comments.
JMMS wishes to thank the School of Mathematical Sciences, Queen
Mary, University of London, for their kind hospitality hosting the EPSRC
visiting fellowship research grant, code GR/N/05550, to which he is also
thankful.  
RV thanks the former Spanish Secretar\'{\i}a de
Estado de Universidades, Investigaci\'on y Desarrollo,
Ministerio de Educaci\'on y Cultura for fellowship no. EX99 52155527,
and also the EPSRC for grant no. MTH 03RAJC6.
RL and JMMS are grateful to the University of the Basque Country
for grant no.  9/UPV 00172.310-14456/2002.
The finantial support given to RL by
the Eusko Jaurlaritza,  under fellowship no. BFI01.412, and by the Spanish 
Ministerio de Ciencia y Tecnolog\'\i a jointly with FEDER funds, under grant
no. BFM2001-0988, is also acknowledged.

\appendix
\section{Appendix}
The 2-form associated with the two Killing vectors
spanning the orbits will
be denoted by
\begin{equation}
\label{eq:w}
\bm w\equiv \xif\wedge\etaf.
\end{equation}
We also define the one-form $\rrf{\xi_A}{\xi_B}{\xi_C}$, relative
to the vector fields $\vec\xi_A,\vec\xi_B,\vec\xi_C$, whose components
are given by
\begin{equation}
  \label{eq:r}
  \rr{\xi_A}{\xi_B}{\xi_C}{\mu} \equiv \xi_{A}^\lambda\,
\xi_B^\alpha\, R_\alpha{}^\rho{}_\lambda{}^\sigma \xi_C^\beta
R_{\beta\rho\sigma\mu},
\end{equation}
where we have introduced the indices $A,B,C=1,2$
to denote the two independent Killing vectors generating the $G_2$ group:
$\xiv_1\equiv\xiv$ and $\xiv_2\equiv\etav$.
It is convenient to define $\ttf{\xi_A}{\xi_B}{\xi_C}$ as
\begin{equation}
  \label{eq:t}
  \ttc{\xi_A}{\xi_B}{\xi_C}{\mu}
\equiv \xi_A^\alpha\eta_{\alpha\nu\rho_1\rho_2}
{\xi_B}_\beta\eta^{\beta\nu\alpha_1\alpha_2}
\xi_C^\lambda
R_{\lambda}{}^{\sigma\rho_1\rho_2}
R_{\alpha_1\alpha_2\sigma\mu},
\end{equation}
where $\eta_{\alpha\beta\mu\nu}$ is the volume element,
so that using (\ref{Bel}), (\ref{eq:r}) and (\ref{eq:t}), a straightforward
calculation allows us to get
{\setlength\arraycolsep{0.1pt}
\begin{eqnarray*}
\CBef{\xi_B}{\xi_A}{}&=& \frac{1}{3}\left\{-\rrf{\xi_B}{\xi_A}{\xi_A}
-2\rrf{\xi_A}{\xi_A}{\xi_B}
-\frac{1}{4}\ttf{\xi_A}{\xi_A}{\xi_B}
-\frac{1}{2}\ttf{\xi_A}{\xi_B}{\xi_A}\right.-\\
&-&\frac{1}{2} \xi_A^\alpha
R_{\alpha\beta\lambda\sigma}~ \xi_A^\rho R_\rho{}^{\beta\lambda\sigma}~
\xif_B -\xi_B^\alpha R_{\alpha\beta\lambda\sigma}~ \xi_A^\rho
R_\rho{}^{\beta\lambda\sigma}~ \xif_A+ \\
&+&\left.\frac{1}{8}(\xiv_A\cdot\xiv_A)
R^{\rho\beta\lambda\sigma}R_{\rho\beta\lambda\sigma}~ \xif_B
+\frac{2}{8}(\xiv_A\cdot\xiv_B)
R^{\rho\beta\lambda\sigma}R_{\rho\beta\lambda\sigma}~ \xif_A\right\}.
\end{eqnarray*}}
Clearly the exterior product $\bm j\wedge \bm w$ contains
only terms of the form $\bm r\wedge \bm w$ and
$\bm t\wedge \bm w$.
We define now the 1-forms $\bm \Sigma_{A}$ 
and $\bm \Omega_{A}$ by
\begin{eqnarray*}
&&\Sigma_{A}{}_{\mu}\equiv\xi^\alpha\eta^\beta\, \xi_A^\lambda\,
R_{\alpha\beta\lambda\mu},\\
&&\Omega_{A}{}_{\mu}\equiv \teta^{\alpha\beta}\, \xi_A^\lambda\,
R_{\alpha\beta\lambda\mu}.
\end{eqnarray*}
Given that $\vec\xi_A$ 
are Killing vectors, we have
$$
\xi^\sigma\eta^\rho R_{\sigma\rho\alpha\beta}=
\nabla_\alpha[\etav,\xiv]_\beta+
2\nabla_\rho\eta_{[\alpha}\nabla^\rho\xi_{\alpha]}.
$$
Contracting this identity with a third Killing vector, and
making the exterior product with $\bm w$, it is straighforward to show
\begin{equation}
  \label{eq:Sigma}
\Sigma_{A}{}_{[\alpha} w_{\mu\nu]}=2\xi_A^\beta\nabla^\rho \eta_\beta\,
w^{ }_{[\mu\nu}\xi_{\alpha;\rho]}
-2\xi_A^\beta\nabla^\rho \xi_\beta\,
w^{ }_{[\mu\nu}\eta_{\alpha;\rho]}
+\xi_A^\beta\nabla_\beta[\etav,\xiv]_{[\alpha} w_{\mu\nu]}
+\frac{1}{3!}\,\xiv_A\cdot [\etav,\xiv]\,\d \bm w_{\alpha\mu\nu}
\end{equation}
Using (\ref{eq:basic}) it is easy to show that
\begin{eqnarray}
  \label{eq:Omega}
\Omega_{A}{}_{[\alpha} w_{\mu\nu]}&=&
\frac{1}{3!}\left(\d*(\bm w\wedge\d\xif_A)\wedge \bm w
-*(\bm w\wedge\d\xif_A)\,\d \bm w\right)_{\alpha\mu\nu}
\nonumber\\
&&-2*(\etaf\wedge\d \xif_A)^\rho w_{[\rho\alpha}\xi_{\nu;\mu]}
+2*(\xif\wedge\d \xif_A)^\rho w_{[\rho\alpha}\eta_{\nu;\mu]}\,.
\end{eqnarray}
Bearing in mind these identities,
the following definitions will be useful later:
\begin{eqnarray}
\label{eq:tSigma}
  &&\tilde{\Sigma}_{A}{}_\alpha\equiv\frac{1}{4}
  \Sigma_{A}{}_{[\alpha} w_{\mu\nu]}w^{\mu\nu}=
  -\frac{1}{3!}\Sigma_{A}{}_\sigma\teta^{\sigma\epsilon}
  \teta_{\alpha\epsilon},\\
\label{eq:tOmega}
  &&\tilde{\Omega}_{A}{}_\alpha\equiv\frac{1}{4}
  \Omega_{A}{}_{[\alpha} w_{\mu\nu]}w^{\mu\nu}=
  -\frac{1}{3!}\Omega_{A}{}_\sigma\teta^{\sigma\epsilon}
  \teta_{\alpha\epsilon}.
\end{eqnarray}
In order to obtain an expression for $\bm r\wedge \bm w$ in terms
of the above quantities we will use the decomposition
of the identity $\delta^\alpha_\beta=
P_\perp^\alpha{}_\beta+P_\parallel^\alpha{}_\beta$ into its part on
the orbit and the corresponding orthogonal element at each point, i.e.
\begin{equation}
\label{eq:det}
P_\perp^\alpha{}_\beta=-\frac{1}{\det}\teta^{\alpha\rho}\teta_{\beta\rho},~~~
P_\parallel^\alpha{}_\beta=\frac{1}{\det}w^{\alpha\rho}w_{\beta\rho},~~~
\det=2w^{\alpha\beta}w_{\alpha\beta}=(\xiv\cdot\xiv)(\etav\cdot\etav)
-(\xiv\cdot\etav)^2.
\end{equation}
Furthermore, it is convenient to compute first $* (\bm r\wedge \bm w)$
and apply the Hodge dual to the final identity.
One can then write
$$
*(\rrf{\xi_A}{\xi_B}{\xi_C}\wedge\bm w)^\gamma=\xi_{A}^\lambda\,
\xi_B^\mu\, R_\mu{}^\rho{}_\lambda{}^\sigma
(P_\perp^\tau{}_\rho+P_\parallel^\tau{}_\rho)
(P_\perp^\epsilon{}_\sigma+P_\parallel^\epsilon{}_\sigma)
\xi_C^\beta
R_{\beta\tau\epsilon\alpha}\teta^{\alpha\gamma}.
$$
Taking into account that
$P_\perp^\epsilon{}_\rho z_{\epsilon\alpha}\teta^{\alpha\gamma}=
\frac{1}{2}P_\perp^\gamma{}_\rho z_{\epsilon\alpha}\teta^{\alpha\epsilon}$
for any 2-form $\bm z$, this identity can be arranged onto its
dual form as follows
\begin{eqnarray}
&&  \rr{\xi_A}{\xi_B}{\xi_C}{[\alpha} w_{\mu\nu]}=
\frac{1}{\det^2}\left\{-6 R_B{}^\rho{}_A{}^\sigma
\tilde\Omega_{C}{}_\rho\teta_{\sigma[\alpha}w_{\mu\nu]}\right.\nonumber\\
&&~~~~-(\xiv_B\cdot\xiv_C)\teta^{\lambda\sigma}
R_{12\lambda\sigma}(*\bm{\tilde \Sigma}_{A})_{\alpha\mu\nu}
-6\,\xi_A{}_\gamma w^{\epsilon\gamma}
\tilde\Sigma_{B}{}_\delta R_{C}{}^\delta{}_{\epsilon[\alpha}
w_{\mu\nu]}\nonumber\\
&&~~\left.\frac{ }{ }-(\xiv_B\cdot\xiv_C)R_{1212}\,
\xi_A^\lambda\left(\eta_\lambda\Sigma_{1[\alpha}w_{\mu\nu]}
-\xi_\lambda \Sigma_{2[\alpha}w_{\mu\nu]}\right)
\right\},
\label{eq:rABC}
\end{eqnarray}
where we have used the indices $A,B,C,1,2$ in some places
also to denote contractions
with $\xiv_A,\xiv_B,\xiv_{C},\xiv,\etav$ respectively, in an obvious manner.

Let us proceed now with $\bm t\wedge \bm w$. 
Using the previous decomposition of $\delta^\alpha_\beta$ for the
summation index $\nu$ in (\ref{eq:t}) and arranging terms conveniently,
we get
\begin{eqnarray*}
  \ttc{\xi_A}{\xi_B}{\xi_C}{[\alpha} w_{\mu\nu]}&=&\frac{1}{\det}
  \left\{-4\,\xi_A^\lambda w_{\rho_2\lambda}\xi_B{}_\beta w^{\alpha_2\beta}
  R_C{}^{\sigma\rho_1\rho_2}R_{\rho_1\alpha_2\sigma[\alpha}w_{\mu\nu]}\right.\\
&&\left.+(\xiv_A\cdot\xiv_B)w^{\rho_1\rho_2}\xi_C^\lambda
R_{\rho_1\rho_2\lambda}{}^\sigma w^{\alpha_1\alpha_2}
R_{\alpha_1\alpha_2\sigma[\alpha}w_{\mu\nu]}
\right.\\
&&\left.+(\xiv_A\cdot\xiv_B)\teta^{\rho_1\rho_2}\xi_C
R_{\rho_1\rho_2\lambda}{}^{\sigma}\teta^{\alpha_1\alpha_2}
R_{\alpha_1\alpha_2\sigma[\alpha}w_{\mu\nu]}
\right\}.
\end{eqnarray*}
The first term has the form $\rr{\xi_C}{v_A}{v_B}{[\alpha}w_{\mu\nu]}$
with $v_A{}^\rho= \xi_A{}_\lambda w^{\rho\lambda}$, and hence
can be expanded using (\ref{eq:rABC}).
For the second and third terms we still have to apply
the same procedure as above to the summation index $\sigma$.
After some calculation we can obtain the following expression
\begin{eqnarray}
\frac{1}{4}\,\ttc{\xi_A}{\xi_B}{\xi_C}{[\alpha} w_{\mu\nu]}&=&
\frac{3!}{\det^3}\left[\xi_B^\lambda
(\eta_\lambda\tilde\Omega_{1}{}_\rho-\xi_\lambda\tilde\Omega_{2}{}_\rho)
\xi_A{}_\gamma w^{\tau\gamma} \xi_C^\epsilon
R_\tau{}^\rho{}_\epsilon{}^\sigma
\teta_{\sigma[\alpha}w_{\mu\nu]}\right.\nonumber\\
&&\left.+\xi_{A}^\lambda
(\eta_\lambda\tilde\Sigma_{1}{}_\rho-\xi_\lambda\tilde\Sigma_{2}{}_\rho)
\xi_B{}_\gamma w^{\tau\gamma} \xi_C{}_\delta w^{\epsilon\delta}
R_\tau{}^\rho{}_{\epsilon[\alpha} w_{\mu\nu]}\right]+\nonumber\\
&&+\frac{1}{4 \det^2}(\xiv_A\cdot\xiv_B)
\left\{3!(2\tilde\Sigma_C{}^\tau w^{\lambda\sigma}+
\tilde\Omega_C{}^\tau \teta^{\lambda\sigma})
R_{\lambda\sigma\tau[\alpha} w_{\mu\nu]}\right.\nonumber\\
&&\left.+[4 (* \tilde\Sigma_C)_{\alpha\mu\nu}
-\xi_C^\epsilon(\eta_\epsilon\Sigma_{1}{}_{[\alpha} w_{\mu\nu]}
-\xi_\epsilon\Sigma_{2}{}_{[\alpha} w_{\mu\nu]})]
\teta^{\lambda\sigma} R_{12\,\lambda\sigma}
\right\}.
\label{eq:tABC}
\end{eqnarray}
Using (\ref{eq:rABC}) and (\ref{eq:tABC}) we are now ready
to obtain the expressions for
\begin{eqnarray*}
\CBef{\xi_B}{\xi_A}{}\wedge \bm w&=&-\frac{1}{3}
\left[\rrf{\xi_B}{\xi_A}{\xi_A}\wedge \bm w+
2\rrf{\xi_A}{\xi_A}{\xi_B}\wedge \bm w\right.\\
&&\hspace{1cm}\left.+\frac{1}{4}\,\ttf{\xi_A}{\xi_A}{\xi_B}\wedge \bm w+
\frac{1}{2}\,\ttf{\xi_A}{\xi_B}{\xi_A}\wedge \bm w\right],
\end{eqnarray*}
which, after using $\xi_A{}_\gamma w^{\lambda\gamma}
(\eta_\lambda\tilde\Omega_{1}{}_\rho-\xi_\lambda\tilde\Omega_{2}{}_\rho)
=-\det \tilde\Omega_A{}_\rho$ and analogous identities, can be
explicitly written as (\ref{eq:collons}).

\end{document}